\documentstyle[stwol,aps,prl,twocolumn,psfig]{revtex}

\def\Journal#1#2#3#4{{#1} {\bf #2}, #3 (#4)}

\def\PLB{{\em Phys. Lett.}  B}

\def\be{\begin{equation}}
\def\ee{\end{equation}}
\def\bea{\begin{eqnarray}}
\def\eea{\end{eqnarray}}

\begin{document}

\title{CAN STRONG QCD IN THE EARLY UNIVERSE RAISE\\ 
THE AXION DECAY CONSTANT?}

\author{JIHN E. KIM}

\address{ Department of Physics and Center for Theoretical Physics\\
Seoul National Univeristy, Seoul 151-742, Korea}

\twocolumn[\maketitle\abstracts{
We report \cite{ckk} that the hypothesis that the upper bound on the 
axion decay constant can be moved up beyond $10^{12}$ GeV
in models with a stronger QCD in the early universe is not 
realized.  This proof is possible by studying the superpotential
in the dual model and obtaining the form of the axion
potential respecting the original global symmetries.  
}]

\section{Introduction}

So far it is known that the cosmological upper bound of the
axion decay constant \cite{bound}
\begin{equation}
F_a\simeq 10^{12}\ {\rm GeV}
\end{equation}
does not allow the Peccei-Quinn symmetry breaking scale at the
grand unification or string scale. Ever since the existence
of this upper bound is known, the axion model encounter either
it is unattractive because it cannot accomodate the GUT scale or
it predicts an intermediate scale. Furthermore in string
models with dilaton $S$, the axion scale is expected to be of
order Planck scale.   With supersymmetry, the QCD
coupling at the string scale is given by the vacuum expectation 
value of the dilaton field $S$, viz.
\begin{equation}
{1\over 32\pi^2}\int d^2\theta f(S)W^aW^a
\end{equation}
where $f$ is a holomorphic fuction of the dilaton superfield $S$,
$S=(8\pi^2/g^2)+i\theta_{QCD}$, and $W^a$ is the gluino
superfield.  In the early universe there is a possibility that 
$\langle S\rangle$ is small, i.e. QCD becomes STRONG even though 
the QCD coupling at low energy is small with $\alpha_c\sim 0.12$.

A naive guess for strong QCD is {\it strong instanton effect}
and a large axion mass; thus settling $\theta=0$ in the early
universe \cite{dvali}.  If so, the initial value of $\langle a
\rangle$ is small compared to $F_a$, and the axion decay constant
bound obtained from the energy density of the universe can be
raised to the GUT or Planck scale, since the coherent oscillation of the
classical axion field starts from almost the minimum point of
$V$. Actually, the conditions for raising $F_a$ to GUT scale are
\cite{ckk}
\begin{equation}
m_a\ge H
\end{equation}
and
\begin{equation}
\left|{\langle a\rangle\over F_a}\right|\le 10^{-3}\sim 10^{-2}
\end{equation}
where $H$ is the Hubble parameter and $\theta_{eff}$ is the low
energy value of the QCD vacuum angle.

In this talk, we restrict the discussion for the case
\begin{equation} 
\Lambda_{QCD}\gg m_{\rm soft}\sim 100\ {\rm GeV}.
\end{equation}

The axion potential in the low energy QCD is obtained under 
the environment of \cite{axion}
\begin{eqnarray}
&m_q \sim 5-10\ {\rm MeV}\nonumber\\
&\Lambda_{\rm QCD}\sim 150\ {\rm MeV}\\
&\langle\bar uu\rangle=\langle \bar dd\rangle=
\langle\bar ss\rangle\sim O(300\ \rm MeV).\nonumber
\end{eqnarray} 
In supersymmetric QCD (SQCD) at high energy, 
we have to check whether what of these
are modified.  Of course, we anticipate an environment of
$\Lambda_{\rm SQCD}$ close to $M_{P}=2.44\times 10^{18}$ GeV.
What about $m_q$?  The value of $m_q$ is related to 
vacuum expectation value of Higgs doublet fields, $H_u$ and $H_d$,
which can be of Planck scale in chaotic inflationary scenario.
Another relevant questions are, $\lq\lq$Do quarks and gluinos
condense?" and $\lq\lq$What is the height of the potential $V$"

\section{Strong SQCD}

In this section, we obtain the axion potential in SQCD.  The
supersymmetric standard model has $N_c=3$ and $N_f=6$. The
superpotential is given by
\begin{equation}
W\ =\ \lambda_uH_uQu^c+\lambda_dH_dQd^c.
\end{equation}
It is known that for vanishing $W$ the quantum moduli space
of degenerate vacua for SQCD ($N_c=3,N_f=6$) is the same as
the classical one \cite{int}. This vacuum degeneracy is lifted 
by the superpotential.  To study the effect of superpotential,
we note the duality of \cite{int}
\begin{eqnarray*}
&SU(N_c)\ with\ N_f\ flavors\\
&\Updownarrow \ \ dual\\
& SU(N_f-N_c)\ with\ N_f\ dual\ quarks.
\end{eqnarray*}
The superpotential of the dual model contains the following
superpotential
\begin{eqnarray}
W_D&=\Lambda_{\rm QCD}(\lambda_uH_uT_u+\lambda_dH_dT_d)\nonumber\\
&+T_uQ_Du_D^c+T_dQ_Dd_D^c
\end{eqnarray}
where subscripts $D$ denote the dual and
\begin{equation}
T_u={Qu^c\over\Lambda_{\rm QCD}},\ T_d={Qd^c\over\Lambda_{\rm QCD}}
\end{equation}
denote composite meson fields made of squarks.
Then the soft terms in the Lagrangian contains
\begin{equation}
-{\cal L}^{(D)}_{\rm soft}\ =\ AW_D+\sum_{I} m_I^2|\phi_I|^2
\end{equation}
from which we note that {\it $H_u,H_d, \{\lambda_uT_u,\lambda_dT_d\}$
have masses of order $m^2$} which is of order
$\Lambda^2_{\rm QCD}$.  We also note that the original squarks with 
positive mass squared have positive mass squared in the dual theory.
Since mesons are bound states of original squarks, they do not have
vacuum expectation values.  Dual squarks can be obtained by
dissociating the scalar baryons (containing $N_c$ original squarks)
into $(N_f-N_c)$ pieces; dual squarks do not have vacuum expectation
values.  Therefore, we conclude
\begin{eqnarray}
&\langle\tilde q\tilde q^c\rangle \propto \langle T\rangle =0\nonumber\\
&\langle qq^c\rangle \propto \langle F_T\rangle =0\\
&\langle \lambda\lambda\rangle \propto \langle T\rangle^{N_f/(N_f-N_c)}=0.
\nonumber
\end{eqnarray}
Therefore, a possible complication, if these condensates are present,
is absent.  Since we are convinced that the condensations are not
present which is manifest in the dual theory as we have shown above, 
we go back to the original theory.  

\section{Axion Potential in MSSM}

In this talk, we concentrate on the minimal supersymmetric standard
model (MSSM).  To obtain the axion potential we include the
$\mu$ term also,
\begin{equation}
W\rightarrow W+\mu H_uH_d
\end{equation}
from which we obtain the soft terms
\begin{eqnarray}
&-{\cal L}_{\rm soft}={1\over 2}m_{1/2}\lambda\lambda+A(\lambda_uH_u
\tilde Q\tilde u^c+\lambda_dH_d\tilde Q\tilde d^c)\nonumber\\
&+B\mu H_uH_d+{1\over 2}\sum_{I}m_I^2|\phi_I|^2+h.c.
\end{eqnarray}
The theory has the following global symmetry 
\begin{eqnarray}
G_{MSSM}=&SU(3)_Q\times SU(3)_{u^c}\times SU(3)_{d^c}\nonumber\\
&\times U(1)_A\times U(1)_X\times U(1)_R
\end{eqnarray}
if we assign appropriate transformation properties for the coupling
parameters.  For the gobal nonabelian transformation under
$SU(3)_Q\times SU(3)_{u^c}\times SU(3)_{d^c}$, $\lambda$'s transform as
\begin{equation}
\lambda_u\sim (\bar 3,3,1)\ ,\ \ \lambda_d\sim (\bar 3,1,3).
\end{equation}
To introduce soft supersymmetry breaking systematically, let us 
introduce spurion superfields
\begin{eqnarray}
&\eta=(1+m_I^2\theta^2\bar\theta^2)\nonumber\\
&Y=(1+16\pi^2m_{1/2}\theta^2)\tau\nonumber\\
&Z_{u,d}=(1+A\theta^2)\lambda_{u,d}\\
&Z_\mu=(1+B\theta^2)\mu\nonumber
\end{eqnarray}
where $\tau=(8\pi^2/g^2)+i\theta_{\rm QCD}$.  From the spurion
superfields, both supersymmetric couplings and soft terms are
given systematically.
To have the $G_{MSSM}$ symmetry the $U(1)$ global charges of the
coupling parameters are assigned as given in Table 1.

\begin{table}[ht]
\centering
\caption{Quantum numbers of superfields and spurions in MSSM}
\vspace{5mm}
\begin{tabular}{|c|c|c|c|}
\hline\hline
 & $U(1)_A$ & $U(1)_X$ & $U(1)_R$ \\
\hline\hline
$Q$          & 1    & 0    & 1    \\ \hline
$u^c$, $d^c$ & 1    & $-1$ & 1    \\ \hline
$H_u$, $H_d$ & 0    & 1    & 0    \\ \hline
$e^{-Y}$     & 12   & $-6$ & 6    \\ \hline
$Z_u$, $Z_d$ & $-2$ & 0    & 0    \\ \hline
$Z_{\mu}$    & 0    & $-2$ & 2    \\ \hline
$d^2\theta$  & 0    & 0    & $-2$ \\ \hline
\end{tabular}
\label{table:1}
\end{table}

Let us introduce a superfield $A$
\begin{equation}
A={1\over F_a}(s+ia+a\theta+f_a\theta^2)
\end{equation}
which is interpreted as a fluctuation of $Y$, i.e. under axion
shift
\begin{equation}
Y\rightarrow Y+A.
\end{equation}
The effective Lagrangian of spurions is obtained from
\begin{eqnarray}
&\int d^2\theta d^2\bar\theta K_{\rm eff}(Y,Y^*,Z,Z^*,\eta)\nonumber\\
&+\int d^2\theta W_{\rm eff}(Y,Z)+h.c.
\end{eqnarray}
For $\langle\lambda\lambda\rangle\sim e^{-i\theta_{QCD}/N}$, then
$e^{-Y/N}$ is present in the effective Lagrangian.  But we have shown that 
the unique ground state which preserves the chiral symmetries has
no branch cut.  Thus instantons would induce a term of the form
\begin{equation}
e^{-nY}\omega(Z)
\end{equation}
in $W_{\rm eff}$.  But selection rules of $G_{MSSM}$ does not
allow any holomorphic $\omega$ which is finite at $\mu\rightarrow 0$.
Therefore, we conclude that the axion potential arises from the first term
of Eq. (19), i.e. from the effective K$\ddot {\rm a}$hler function $K_{\rm
eff}$. 

For $n=1$,
\begin{equation}
K_{\rm eff}\propto e^{-Y}{\rm Det}(Z_uZ_d)Z_\mu^{*3}F(\eta)+h.c.
\end{equation}
where $F(\eta)$ is an arbitrary function of $\eta$.
Note that $K_{\rm eff}$ is invariant under $U(1)_A\times U(1)_X\times
U(1)_R$.  To obtain the axion potential, one must insert
$D_\eta$, or $F_{Y,Z}$ and $F_{Y,Z}^*$ insertions.  

 From Eq. (21), we represent the order of soft parameters as
\begin{equation}
[m_{\rm soft}]^2=\{m_i^2,AB^*,16\pi^2m_{1/2}B^*\}.
\end{equation}
Since instantons give dominant contributions for 
$\rho\sim \Lambda^{-1}_{QCD}$, the axion potential is estimated
as
\begin{eqnarray}
&V_a\ \simeq\ e^{ia/F_a}\left({1\over 16\pi^2}\right)^6
\mu^{*3}{\rm Det}(\lambda_u\lambda_d)\nonumber\\
&[m_{\rm soft}]^2
\Lambda^{-1}_{QCD}+h.c.
\end{eqnarray}
A careful study of this axion potential is given in Ref. [1].
Actually, one instanton diagram giving Eq. (23) can be found as
shown in Fig. 1. Twelve quark lines and six gaugino lines (corresponding
to twice the index of adjoint representation of $SU(3)_{\rm color}$)
are coming out from the instanton vertex.  Quark lines have Yukawa 
couplings to $H_{u,d}$ and gaugino and quark lines have gauge couplings
to squarks.  Two gauginos have soft gaugino mass coupling,
two Higgs doublets have the $B\mu$ term coupling, and two squarks and a
Higgs doublet have the $A$ term coupling; thus leading to
Eq. (23). In Eq. (23), only $[m_{\rm soft}]^2$ is given, but from
Fig. 1 we can see the explicit dependence. In Fig. 1, the $\mu^*
\lambda_{u,d}$ vertices arise from $|\partial W/\partial H_{u,d}|^2$ 
term.  Thus from Fig. 1,
we obtain
\begin{eqnarray}
&(\theta_{\rm eff})_{\rm MSSM}=\theta_{QCD}+{\rm Arg} ({\rm Det}\lambda_u
\lambda_d)\nonumber\\
&+3{\rm Arg}m_{1/2} -3{\rm Arg}(\mu B)
\end{eqnarray}
where $\theta_{QCD}=\langle a\rangle/F_a$.

\begin{figure*}
\begin{center}
\mbox{
\psfig{figure=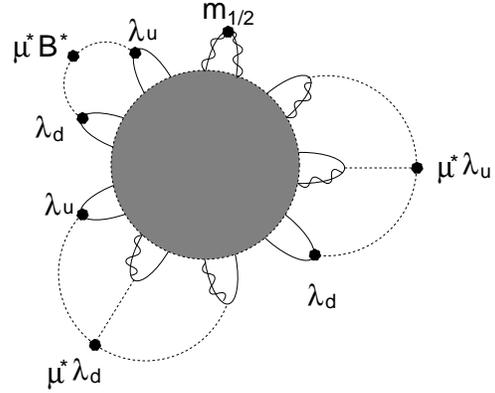,height=2.0in}
      }
\end{center}
\caption{Instanton graph for the axion potential Eq. (22) of the
MSSM. The solid lines with and without waves around the instanton
denote the gluino and the quark modes, respectively, while dotted
lines are Higgs and squark fields.  The dark blobs represent
the insertions of complex couplings. 
\label{fig:instanton}}
\end{figure*}
\section{In Early Universe}

The early universe values of couplings
\begin{equation}
\lambda_{u\ \rm in},\lambda_{d\ \rm in},\mu_{\rm in},A_{\rm in},
B_{\rm in},m_{1/2\ \rm in}
\end{equation}
and present values of couplings
\begin{equation}
\lambda_{u\ \rm eff},\lambda_{d\ \rm eff},\mu_{\rm eff},A_{\rm eff},
B_{\rm eff},m_{1/2\ \rm eff}
\end{equation}
should be almost the same for the conditions (3) and (4) to
be satisfied.  Then we expect
\begin{equation}
\theta_{\rm eff}\ \simeq\ \theta_{\rm in}.
\end{equation}
Referring Eq. (24), we define $\delta\theta$ as
the value of $\theta$ except $\langle a\rangle/F_a$.
If Eqs. (3) and (4) are satisfied,
the strong QCD in the early universe determines the present
value $\theta_{\rm eff}$ at almost 0 and then the axion energy
crisis does not occur even if $F_a\sim M_P$.

Certainly $\theta_{\rm eff}$ is the value where $V_a$ is the
minimum.  If $\theta_{\rm in}$ is also the value where $V_a$
is the minimum, then there is a possibility that Eq. (27) is
satisfied.  However, for this to happen the axion potential 
must be steep enough so that the minimum of the axion
potential is quickly achieved in the early universe, i.e.
$(\langle a\rangle/F_a)_{\rm in}=0$ and $\delta\theta\simeq 0$.
However, the axion potential is sufficiently suppressed by
$m_{\rm soft}$ and hence $\theta_{\rm in}$ does not reach 
$\theta_{\rm eff}$ in the early universe.

Our question is, {\rm if $\Lambda_{QCD}\gg m_{\rm soft}$,
then can $F_a\gg 4\times 10^{12}\ {\rm GeV}$ ? For $a$ to roll
down the hill in the early universe,} 
\begin{equation}
m_a\ge H
\end{equation}
where $H$ is the Hubble parameter in the early universe.
 As we have seen above, the axion potential is sufficiently
suppressed in MSSM
\begin{equation}
\mu^{*3}[m_{\rm soft}]^2.
\end{equation}
To raise $F_a$, we need
\begin{equation}
\delta\theta_{\rm in}\ =\ {\langle a\rangle\over F_a}
-\theta_{\rm eff}\le 10^{-2}-10^{-3}.
\end{equation}
We estimate 
\begin{eqnarray}
&\left({m_a\over H}\right)_{MSSM}
\simeq 10^{-11}C^{-1/2}\left({4\times 10^{12}\over F_a}\right)\nonumber\\
&\left({10^5\over M_m}\right)\left({\mu\over 10^2}\right)^{3/2}
\left({10^2\over \Lambda_{QCD}}\right)^{1/2}
\end{eqnarray}
where $M_m$ is the messenger scale for the supersymmetry breaking
and every mass is in GeV units.  Therefore, we conclude
$m_a\ll H$.  An extension of the MSSM to the next minimal
supersymmetric standard model (NMSSM) does not improve the
result \cite{ckk}.

\section{Conclusion}

We have shown that the strong QCD in the early universe cannot 
raise the cosmological upper bound on the axion decay constant
$F_a$ in the MSSM and NMSSM.  We also have
shown the method to calculate the form of the axion potential 
in supersymmetric models from the symmetry argument.

\section*{Acknowledgments}

This work is supported in part by KOSEF through CTP of Seoul
National University, Ministry of Education BSRI-96-2418, and SNU-Nagoya
Collaboration Program of Korea Research Foundation.

\end{document}